# Title:

# Do They Accept or Resist Cybersecurity Measures?

# Development and Validation of the 13-Item Security Attitude

# Inventory (SA-13)[1]

Authors:

Cori Faklaris, Laura Dabbish and Jason I. Hong, *Carnegie Mellon University*

Corresponding author:

Cori Faklaris, cfaklari@cs.cmu.edu, Human-Computer Interaction Institute, Carnegie Mellon University, 5000 Forbes Ave., Pittsburgh, PA 15213 (Permanent email: cori@corifaklaris.com)

Funding:

This work was supported by the U.S. National Science Foundation, grant no. CNS-1704087. The first author also received fellowship support from the CyLab Security and Privacy Institute and the Center for Informed Democracy and Social Cybersecurity, both at Carnegie Mellon University. Sponsors were not involved in any phase of research or article preparation.

---

[1] Abbreviations used in this article: SA-13, for 13-item Security Attitude inventory; SA-Engagement, for the Engagement with Security Measures scale; SA-Attentiveness, for Attentiveness to Security Measures scale; SA-Resistance, for Resistance to Security Measures scale; SA-Concernedness, for Concernedness with Improving Compliance scale; RSec, for Recalled Security Actions scale; SeBIS, for Security Behavior Intention Scale.

# Abstract

We present *SA-13,* the 13-item Security Attitude inventory. We develop and validate this assessment of cybersecurity attitudes by conducting an exploratory factor analysis, confirmatory factor analysis, and other tests with data from a U.S. Census-weighted Qualtrics panel ($N$=209). Beyond a core six indicators of *Engagement with Security Measures* (SA-Engagement, three items) and *Attentiveness to Security Measures* (SA-Attentiveness, three items), our SA-13 inventory adds indicators of *Resistance to Security Measures* (SA-Resistance, four items) and *Concernedness with Improving Compliance* (SA-Concernedness, three items). SA-13 and the subscales exhibit desirable psychometric qualities; and higher scores on SA-13 and on the SA-Engagement and SA-Attentiveness subscales are associated with higher scores for security behavior intention and for self-reported recent security behaviors.

SA-13 and the subscales are useful for researchers and security awareness teams who need a lightweight survey measure of user security attitudes. The composite score of the 13 indicators provides a compact measurement of *cybersecurity decisional balance* – the degree to which system users' acceptance of security measures weighs against their resistance in deciding whether to adopt secure behaviors or to fully comply with policies and advice. SA-13 has potential to act as a diagnostic for this decisional balance and to measure the effects of security interventions outside of system log data.

# 1. Introduction

Understanding human behavior is critical for securing computing systems and reducing risk of cybercrime [36]. In 2020, McAfee estimated that the global costs of cybercrime had jumped more than 50% in two years, to over $1 trillion [67]. Their researchers cited the "lack of organization-wide understanding of cyber risks" as why many companies and agencies fall victim to so-called social engineering attacks. In such attacks, malicious actors leverage social norms and other techniques to persuade legitimate users to grant them access to systems that are otherwise secured through technical means [78].

Many users can help prevent or defend against such attacks by complying with expert security advice and by using and not bypassing security tools. These fall into four categories [25]: creating strong and unique passwords for each account; making sure that software is kept up-to-date so that the needed security fixes are applied; staying alert for scams, hoaxes such as phishing, and for "fake news" in email, website browsing and social media; and using multi-factor authentication for accounts where it is offered. However, research in usability and security [1,7,16,22,44,47,62,73–76] shows that compliance is far from universal, due to nonexpert users' mental models about cybersecurity risks and their evaluations of the advice and tools available to address these risks.

To this end, we see an urgent need for better psychometric tools to assess end users' attitudes [6,24] about cybersecurity. Such measurements, easy to deploy across an organization in an emailed survey or put in front of individual users via paper or internet quiz, can help information security professionals to determine which users are primed for security awareness training, or which users are resistant to security compliance. For researchers, psychometric tools can help to assess cybersecurity attitudes in a given population sample and to quantify the effects on security attitudes of introducing new security tools or risk awareness campaigns. Such measures can also help researchers to identify the degree to which this attitude measure, and/or its factors, help to explain and predict people's intentions to put expert security advice into action, and their instances and frequency of acting.

The current state-of-the-art psychometric tools for IT professionals — the 31-item *Personal Data Attitude* measure for adaptive cybersecurity [2], and the 63-item *Human Aspects of Information Security Questionnaire*, or HAIS-Q [53] — seem inadequate for general use in end-user security research, as they are specific to personal data attitudes and to information security awareness respectively, and their length could lead respondents to abandon the survey. Another measure more suited to general use, our *SA-6*, for six-item security attitude scale [30], is a brief, yet reliable and valid measure of the degree to which people have favorably evaluated security measures, but it lacks items to measure the degree of ambivalent, indifferent, or unfavorable evaluation that weighs in the balance of their decision-making. Such *decisional balance* [45] long has shown importance in changing ingrained behaviors that impact health, such as quitting smoking [71]; and, more recently, in explaining reasoning about cybersecurity, such as why people fail to correctly identify phishing emails and make sense of browser warnings [23], or how they weigh the costs and benefits of following security advice [28,61]. While not invoking decisional balance explicitly, several studies of users' mental models have noted users' noncompliance with security advice due to its perceived negative impact on themselves or their social ties, or on their self- or social esteem, or a combination of these [16,35,62,74]. A multi-factor measurement model of security attitudes that includes items about security noncompliance would give researchers and industry professionals a choice to assess separate aspects of security attitudes that weigh in users' decisional balance, or to quantify this balance in a single score.

## 1.1. Why understanding end users is important to cybersecurity

Computer users have long been called the "weakest link" [1,62,75] in ensuring that confidential information shared in a computer system is kept safe, and that the integrity of shared information is likewise protected. Their data "keys" — identification (that the user is "on the list" to be let in) and authentication (that they are who they say they are) — can be stolen from the end users or the system administrators if they, in effect, leave these keys lying around in an unprotected machine or area of cyberspace.

Because of this "weakest link," users' *mental models* – the thoughts and feelings underpinning their decisions to follow, or not follow, basic cybersecurity advice such as creating hard-to-guess passwords and not reusing them — are a chief concern of human-computer interaction, particularly the subfield of usable privacy and security, and of human-factors specialists in information science, computer science and engineering, and cybersecurity. Borrowing from cognitive psychology [23], many researchers in these fields have employed empirical methods such as in-person interviews or surveys to document these mental states in everyday use contexts. Notable studies include Wash's investigation of how home computer users conceptualize possible threats such as hackers [74], Ion et al.'s comparison of how security experts vs. nonexperts prioritize practices such as changing passwords or installing software updates [44], and Kang et al.'s use of participant drawings to document how they visualize their data moving on the internet [47]. More recent work centered on smartphones has proposed new or extended models of security compliance [8,9,72], examined why those users do or do not install and use a password manager [7] or generating complex passwords with such an app [76], or why smartphone users fall victim to phishing attacks [73]. Thompson et al. looked at how psychological ownership and other concepts impacted behavioral intentions and behaviors with both home computers and mobile devices [69].

Such user studies are resource-intensive, both in time to conduct the studies and to develop the interview scripts and the survey instruments. An appealing alternative is to adapt or re-use an already published, *lightweight* measure of users' mental states – one short enough to be likely to not be abandoned mid-survey, and to be combined with other questions such as about participant demographics without making themselves too long. The best psychometric surveys are shown through statistical analysis to be reliable and valid as measurements of the concepts they purport to measure, but to also not be so specific as to not be re-usable by others.

## 1.2. 'Security sensitivity' as a basis for measuring security attitudes

A particular body of work on understanding users' attitudes about cybersecurity as a means toward "transforming the weakest link" [62] has developed around the concept of *security sensitivity* [30]. Das et al. describe security sensitivity as comprised of users' awareness of both the relevant threats and the means to combat them, along with motivation to comply with security advice and knowledge of how to use tools that will protect systems across the technology stack [15,16]. These studies found that such compliance with advice and conformity with security practices are driven in part by social influences [16,18] in the form of social proof via Facebook posts [17,18], conversations sparked by personally experiencing a privacy or security breach [16] and hearing or seeing news about privacy and security breaches [19].

Building particularly on this work with security sensitivity, we developed and validated a relatively short measure of security attitudes called SA-6 [30]. This is comprised of six statements that, overall, describe the degree to which a person has already decided to accept and adopt security advice, to be rated on a bidirectional agreement scale (5=Strongly Agree to 1=Strongly Disagree) and then averaged to generate the final score. The item wordings are drawn from Das et al. and similar research, such as "I am extremely knowledgeable about all the steps needed to keep my online data and accounts safe" and "I often am interested in articles about security threats." We also developed a second measure to be used as an outcome variable in testing SA-6 that we named Recalled Security Actions, or RSec for short [30]. Using this and the separately developed Security Behavior Intention Scale, or SeBIS [25] and other measures such as the Internet Know-How scale [47], we found that SA-6 exhibited predictive validity and other desirable psychometric properties for an attitude scale.

SA-6 is much shorter and more suitable for use in the general population than two other recently developed measures of security awareness, the 31-item Personal Data Attitude measure for adaptive cybersecurity [2], and the 63-item Human Aspects of Information Security Questionnaire, or HAIS-Q [53]. However, experience has shown us that even six items might be too many to include in a survey protocol or the screen interface for an online study that already is stuffed with other questions. Moreover, we note that some of the item wordings describe engagement as passive, and some describe engagement as more active. It would be useful to re-analyze SA-6 to see if it can be split into more than one factor that indicate a more passive vs. active engagement with security advice, and therefore to more accurately pinpoint the attitudes that predict and explain someone's degree of security compliance.

## 1.3. Decisional balance and security noncompliance

Even if someone has a high degree of awareness, motivation and knowledge for complying with security advice, such as using a Virtual Private Network (VPN) or installing a software update, they still may fail to act. People make such decisions by weighing what they know of the action's utilitarian gains and losses for themselves and for others, along with the degree of self- or social approval – or disapproval — that action carries [45]. This *decisional balance* has been seen as key for changing behaviors that impact health, such as quitting smoking [71]. In cybersecurity, decision strategies have helped explain why people fail to correctly identify phishing emails and make sense of browser warnings [23], or how they weigh the costs and benefits of following security advice [28,61].

While not invoking decisional balance explicitly, several studies of users' mental models have noted users' noncompliance with security advice due to its perceived negative impact on themselves or their social ties, or on their self- or social esteem, or a combination of these. For example, some users of computing devices have remarked that the use of extra cybersecurity measures such as encryption is evidence of "paranoia" [16,35]. Users may feel that they only visit "trusted" websites that won't lead to a data breach, or that they are not rich or important enough to attract a hacker's attention [74]. Further, the rigidity of security requirements can lead users to feel "ambushed" at inopportune times by a security feature demanding new input, such as being required to deal with password policies at login [62]. Most people also see complying with cybersecurity as a secondary goal at best in their use of computing devices, which can lead them to ignoring security advice or cutting corners with requirements [62]. The perceived cost of compliance (the "level of effort or financial cost associated with incorporating a protective measure") is one of five independent factors in the Technology Threat and Avoidance Theory model [36].

We see the need to, at the least, add to our SA-6 security attitude measure to address both negative and positive mental states in which users are conscious of failing to act.

## 1.4. Candidate items for assessing noncompliant attitudes

In our past work on SA-6 [30], we generated an initial set of 200+ candidate scale items, drawn from empirical research by Das et al. [15–19], Egelman and Peer [25] and other work in usable security and in the psychology of computer use [35,74], as well as from their own expertise in the subject. These items were comprised of statements of cybersecurity attitudes rated on a 5-point Likert-type agreement scale (1=Strongly disagree, 5=Strongly agree). Following best practices [21,34,41,52], we recruited experts and nonexperts in security research to review the items for content adequacy and for clarity and unambiguousness of wording, respectively. We then examined survey responses to further whittle the list down to a final 48 candidate items (Appendix A). Responses to these items were collected in samples drawn, first, from the Amazon Mechanical Turk crowdsourcing platform, and second, from the Qualtrics panel recruitment service. We then used these responses to create the SA-6 instrument [30].

For the current study, we re-analyze these two datasets to determine a multi-factor model that can expand the SA-6 measure to include those items that speak to decisional balance with security noncompliance.

## 1.5. Applying the Theory of Reasoned Action to cybersecurity

For our prior work on SA-6 [30], we applied the Theory of Reasoned Action, also known as the Theory of Planned Behavior [3–5,32]. This theory posits that attitude, influenced by background factors, helps to predict and explain behavior intention and actual behavior (Figure 1).

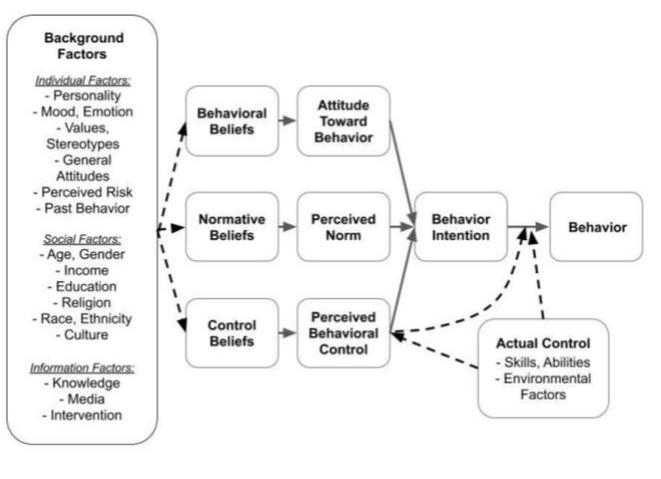

*Figure 1: Diagram of the Theory of Reasoned Action as developed by Ajzen and Fishbein*

Guided by this theory, we collected variables for their study that are thought to relate with or to vary with security attitude, and that previously been found to relate with or vary with security behavior intention [25,26]. Such data collection is considered a best practice [21,34,41,52] in order to test the degree to which these variables are associated with the newly developed scale. These included the following previously validated measures, computed as a composite of several survey items and used to test convergent validity [30]: Security Behavior Intentions [25], Internet Know-How [47], Confidence in Using Computers [33], Need for Cognition [13], Consideration of Future Consequences [68], General Decision-Making Styles [65], Health/Safety Risk Perception and Health/Safety Risk Taking [11], Impulsiveness [59],

Web-oriented Digital Literacy [38], General Self-Efficacy and Social Self-Efficacy [66], Privacy Concerns [12,51] and the "Big Five" personality factor of Extraversion [37].

We also collected variables for age, gender, education, income, frequency of personally experiencing security breaches, frequency of social ties experiencing security breaches, and frequency of hearing or seeing news about security breaches. These were directly collected as single items rated on a 5-point scale and used to test discriminant validity [30].

Finally, as a proxy for security behavior, which is the main outcome variable in the TRA diagram, we created a nine-item measure that we named the Recalled Security Actions inventory (RSec) by rewording the items in the Security Behavior Intentions Scale (SeBIS) to refer to actions that participants recalled performing during the prior week. This was used to test predictive validity [30].

To provide comparisons with this prior work, we include the same measures in the present study to assess convergent, discriminant and predictive validity.

# 2. Method

## 2.1. Datasets

In the present study, we use the two independently collected datasets from our prior SA-6 work [30]. These allow for statistical cross-validation of the initial item selection in the second Qualtrics dataset and for comparison with the earlier paper's results. The datasets comprise various candidate items that we based on reports in prior studies in usable privacy and security, and collected via Likert-type interval scales. Each dataset is comprised of more than 200 observations, an accepted threshold for convenience samples in survey research [21,46].

The first dataset, used solely for identifying 13 items and the viability of a four-factor model of SA-13, is the set of 478 online survey responses collected from the Amazon Mechanical Turk crowdsourcing platform. These are used for initial Item Selection in Phase 1, as described below, with results reported in Appendices A-C.

The second dataset, for validating the resulting SA-13 inventory in a new dataset and confirming the four-factor structure and models, is the set of 209 responses collected from a U.S. Census-weighted Qualtrics panel. The Results section is based in this dataset, starting with the Exploratory Factor Analysis in Phase 2 and continuing with analyses in Phase 3.

# 2.2. Procedure

## 2.2.1. (Phase 1) Selecting Candidate Items for the SA-13 Inventory

Using, first, the 475 online survey responses from our prior SA-6 work [30], we examined their preliminary 48 items and whittled them down to 13 using an iterative process of exploratory factor analysis and internal consistency analysis in SPSS. We used this larger dataset for the item selection process because it provided an approximate 10:1 ratio of item responses to candidate items, which is considered ideal [39,41,52]. In an unrotated Principal Axis Factoring, we used the sum of squares loadings to find that the 13 items, constrained to load onto four factors, explain 54.9% of the dataset variance, which we deemed acceptable [39].

See Appendix A for a table of all 48 candidate items and their direct sources; Appendix B for a table of the means, standard deviations and correlations among the selected 13 items in the $N_{EFA}$=475 dataset; and Appendix C for the unrotated factor loadings and scree plot.

## 2.2.2. (Phase 2) Developing and Refining the SA-13 Inventory

### 2.2.2.1. Exploratory Factor Analysis

Moving onto the 209-response Qualtrics dataset from our prior SA-6 work [30], we repeated the investigation of a one-, two-, three-, and four-factor solution in SPSS. We used the scree plot test [39] to identify four factors as the maximum number of factors that might be meaningful. This scree plot of eigenvalues (Figure 2) reveals a break at Factor 4, with eigenvalues leveling off beyond the break and signaling the unimportance of Factors 5-13.

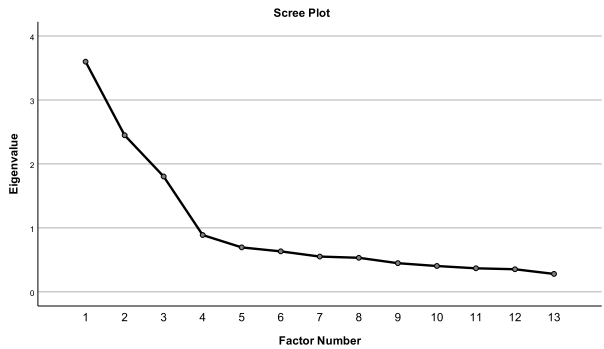

*Figure 2: Scree plot of eigenvalues for the exploratory factor analysis for SA-13 in the Qualtrics dataset (N=209). The break at Factor 4 indicate that a four-factor solution is preferred.*

Next, we examined the factor statistics and the wording of the items that loaded on each factor using Principal Axis Factoring (PAF) and then adding Promax (oblique) rotation, with >=|.40| as the threshold for an item to be considered as loading on a factor [39]. These led us to select the four-factor solution as both the most interpretable (based on wordings) and likely to validate (based on statistics).

## 2.2.2.2. Internal Consistency

We examined the internal consistency of these factor solutions by calculating Cronbach's alpha and alpha-if-item-deleted using SPSS [39]. These aided the iterative development process, by using alpha = .70 as the threshold for inclusion or deletion of items for each factor that was then tested using PAF, looking for a corresponding score for alpha if item deleted that was less than the overall alpha.

## 2.2.3. (Phase 3) Validating the SA-13 Inventory

### 2.2.3.1. Goodness-of-fit

Using the four-factor model developed above, we conducted exploratory structural equation modeling and confirmatory factor analysis in Mplus, following established best practices [39,42,43,48,64]. We used the maximum-likelihood method to estimate parameters. We chose the Comparative Fit Index (CFI) and the Root Mean Square Error of Approximation (RMSEA) as our primary fit indices, with an acceptable CFI value considered to be greater than 0.90 and an acceptable RMSEA value considered to be less than 0.08 [39,43]. We also report the Standardized Root Mean Residual (SRMR), similar to RMSEA but more forgiving of complexity; the Tucker-Lewis Index (TLI), which reflects the fit of the measurement model against the null (baseline) model; and the ratio of the chi-statistic ($x^2$) to degrees of freedom ($df$), which is a better test of fit than $x^2$ alone in larger samples and ideally is below 3.00 [39,43].

Lastly, we conducted several tests of model invariance to confirm SA-13's overall robustness [49,56,63]. For these model tests, we also report the Bayesian Information Criterion (BIC), the Akaike Information Criterion (AIC), and their ratios to degrees of freedom ($df$), with lower values preferred.

### 2.2.3.2. Reliability analyses

As noted above, we conducted reliability analyses for the four SA-13 subscales by calculating Cronbach's alpha and alpha if item deleted using SPSS, with .70 as the threshold for inclusion in this report.

Separately, we calculated Cronbach's alpha for each of the non-SA-13 measures in the dataset (used for validity testing, structural equation modeling and path analyses). We excluded those measures from validity testing that did not exhibit an alpha of .70. A table of reliability statistics for all measures is included in Appendix D.

### 2.2.3.3. Validity testing

Following the methods used in our SA-6 work [30], we tested convergent validity using Spearman correlations in SPSS. These looked for significant associations among the various other variables collected, such as Internet Know-How and Privacy Concerns, and SA-13 and its four subscales.

Similar to the prior work, we also tested discriminant validity using analyses of variance (ANOVAs) in SPSS. These looked for the expected variances in SA-13 and its four subscales by the levels of the various demographic and experience variables collected, such as Breach Experience and Age.

Finally, we tested for predictive validity using linear and stepwise regression, again using SPSS. For these, we carried out several tests of directed relationships with SA-13, its subscales, and SeBIS as the explanatory variables, and RSec as the outcome variable.

# 3. Results

We chose and determined the 13 final items that make up the SA-13 inventory with an iterative exploratory factory analysis (Appendices A-C, Figure 2). The result adds seven indicators of security noncompliance to the initial six items of SA-6 (Table 1, Appendix E). Our initial four-factor model (Table 2) displayed goodness-of-fit: $x^2(32)=43.238$, $p>.05$. We labeled the four scales *Engagement with Security Measures (SA-Engagement* for short), *Attentiveness to Security Measures (SA-Attentiveness), Resistance to Security Measures (SA-Resistance)* and *Concernedness with Improving Compliance (SA-Concernedness).* We found the first three to display good internal consistency (Tables 3-5), supporting their use as independent scales.

Subsequent analysis showed that SA-Engagement displays in particular convergent, discriminant and predictive validity that is close to that of SA-6. This supports its use as a smaller version of SA-6. Furthermore, the overall SA-13 inventory and the SA-Resistance subscale display desirable psychometric qualities throughout testing. This supports the use of SA-13 (Appendix E) as a robust attitude measure that incorporates noncompliance and the use of SA-Resistance to focus on this noncompliance.

## 3.1. Finalized Items of the SA-13 Inventory

Item Correlations for Qualtrics Dataset (N=209, * p<.05, ** p<.01)

| Variable | Item | SAE1 | SAE2 | SAE3 | SAA1 | SAA2 | SAA3 | SAR1 | SAR2 | SAR3 | SAR4 | SAC1 | SAC2 | SAC3 |
|---|---|---|---|---|---|---|---|---|---|---|---|---|---|---|
| SAE1 | I often am interested in articles about security threats. | 1 | | | | | | | | | | | | |
| SAE2 | Generally, I diligently follow a routine about security practices. | .541** | 1 | | | | | | | | | | | |
| SAE3 | I seek out opportunities to learn about security measures that are relevant to me. | .607** | .599** | 1 | | | | | | | | | | |
| SAA1 | I always pay attention to experts' advice about the steps I need to take to keep my online data and accounts safe. | .336** | .405** | .491** | 1 | | | | | | | | | |
| SAA2 | I am extremely knowledgeable about all the steps needed to keep my online data and accounts safe. | .356** | .442** | .441** | .424** | 1 | | | | | | | | |
| SAA3 | I am extremely motivated to take all the steps needed to keep my online data and accounts safe. | .421** | .463** | .463** | .574** | .568** | 1 | | | | | | | |
| SAR1 | I usually will not use security measures if they are inconvenient. | .000 | .132 | .022 | .096 | -.005 | .143* | 1 | | | | | | |
| SAR2 | There are good reasons why I do not take the necessary steps to keep my online data and accounts safe. | -.002 | -.018 | -.095 | -.129 | -.176* | -.056 | .439** | 1 | | | | | |
| SAR3 | I am too busy to put in the effort needed to change my security behaviors. | -.041 | .024 | -.014 | .074 | -.019 | .144* | .423** | .468** | 1 | | | | |
| SAR4 | I have much bigger problems than my risk of a security breach. | .017 | .056 | .029 | .018 | -.110 | .058 | .371** | .460** | .554** | 1 | | | |
| SAC1 | I want to change my security behaviors to improve my protection against threats (i.e. phishing, computer viruses, identity theft, password hacking) that are a danger to my online data and accounts. | .210** | .172* | .304** | .106 | .032 | .180** | .005 | -.087 | -.067 | .017 | 1 | | |
| SAC2 | I worry that I'm not doing enough to protect myself against threats (i.e. phishing, computer viruses, identity theft, password hacking) that are a danger to my online data and accounts. | .064 | -.055 | .093 | -.081 | -.149* | -.084 | -.052 | -.102 | -.265** | -.045 | .391** | 1 | |
| SAC3 | I want to change my security behaviors in order to keep my online data and accounts safe. | .210** | .199** | .205** | .179** | .081 | .231** | .086 | .029 | -.070 | .002 | .553** | .366** | 1 |
| | Mean (1.00-5.00 scale) | 3.50 | 3.74 | 3.53 | 3.79 | 3.44 | 3.80 | 3.12 | 3.22 | 3.52 | 3.18 | 3.72 | 3.37 | 3.72 |
| | Standard Deviation | 1.12 | 0.99 | 1.07 | 0.91 | 1.12 | 1.01 | 1.24 | 1.32 | 1.22 | 1.20 | 1.04 | 1.16 | 1.00 |

*Table 1: Descriptive statistics for the final items chosen for SA-13 show that the original six items (SAE1-3, SAA1-3) are largely not correlated with the final seven items (SAR1-4, SAC1-3), adding to evidence for a factor solution that results in 3-4 subscales.*

Our preliminary work was an iterative process of exploratory factor analyses to identify the items with means most closely associated with each other, the results of which are summarized in Appendices A-C. We settled on adding a final seven items, indicating degrees of noncompliance, to the initial six items of SA-6. See Table 1 for item correlations, means, and standard deviations.

## 3.2. Factor Structure and Internal Consistency

Next, we conducted Principal Axis Factoring, first without and then with Promax (oblique) rotation. This led us to tentatively confirm the four-factor solution as both the most interpretable (based on wordings) and likely to validate (based on statistics), shown in Table 2. The chi-square test of the observed sample distribution with the expected probability distribution was not significant, indicating goodness-of-fit for the four-factor model: $x^2(32)=43.238$, $p>.05$.

| Variable | Item | Factor 1 | 2 | 3 | 4 |
|---|---|---|---|---|---|
| SAR3 | I am too busy to put in the effort needed to change my security behaviors. | 0.738 | | | |
| SAR2 | There are good reasons why I do not take the necessary steps to keep my online data and accounts safe. | 0.703 | | | |
| SAR4 | I have much bigger problems than my risk of a security breach. | 0.696 | | | |
| SAR1 | I usually will not use security measures if they are inconvenient. | 0.583 | | | |
| SAE3 | I seek out opportunities to learn about security measures that are relevant to me. | | 0.753 | | |
| SAE1 | I often am interested in articles about security threats. | | 0.748 | | |
| SAE2 | Generally, I diligently follow a routine about security practices. | | 0.624 | | |
| SAA1 | I am extremely motivated to take all the steps needed to keep my online data and accounts safe. | | | 0.827 | |
| SAA2 | I am extremely knowledgeable about all the steps needed to keep my online data and accounts safe. | | | 0.599 | |
| SAA3 | I always pay attention to experts' advice about the steps I need to take to keep my online data and accounts safe. | | | 0.597 | |
| SAC1 | I want to change my security behaviors in order to keep my online data and accounts safe. | | | | 0.738 |
| SAC2 | I want to change my security behaviors to improve my protection against threats (i.e. phishing, computer viruses, identity theft, password hacking) that are a danger to my online data and accounts. | | | | 0.714 |
| SAC3 | I worry that I'm not doing enough to protect myself against threats (i.e. phishing, computer viruses, identity theft, password hacking) that are a danger to my online data and accounts. | | | | 0.571 |

Extraction method: Principal Axis Factoring. Rotation method: Promax with Kaiser Normalization. Rotation converged in 6 iterations.

*Table 2: Unstandardized loadings for the SA-13 indicators (values >.400), sorted by size. The matrix shows that each indicator loads on just one factor, which is desirable for a clean solution.*

Given the item wordings, we labeled the factors as follows: Factor 1, comprising variables SAR1-4, we named *Resistance to Security Measures (SA-Resistance),* to summarize the anti-compliance attitudes expressed in the individual item wordings; Factor 2, comprising variables SAE1-3, we named *Engagement with Security Measures (SA-Engagement),* to describe the attitudes of proactivity and interest in learning described by the measures; Factor 3, comprising variables SAA1-3, we named *Attentiveness to Security Measures (SA-Attentiveness)*, to describe the attitudes of alertness and motivation to follow advice; and Factor 4, comprising variables SAC1-3, we named *Concernedness with Improving Compliance (SA-Concernedness),* to summarize "feeling one could do better". Factors 2 and 3 correspond to SA-6, the six-item scale previously published using this dataset, and we refer to these first when discussing the four in the rest of the paper.

We found that Cronbach's alpha for three of these factors exceeds the traditional .70 threshold: *SA-Engagement* alpha = .81, *SA-Attentiveness* alpha = .76, and *SA-Resistance* alpha = .77. This indicates good internal consistency and supports their use as independent scales. The *SA-Concernedness* alpha just grazes the threshold, at .69, indicating that its use as an independent scale should be approached with caution. Tables 3-6 show the item-total statistics for each.

| Variable | Item | Corrected Item-Total Correlation | Alpha if Item Deleted |
|---|---|---|---|
| SAE1 | I often am interested in articles about security threats. | 0.643 | 0.748 |
| SAE2 | Generally, I diligently follow a routine about security practices. | 0.635 | 0.755 |
| SAE3 | I seek out opportunities to learn about security measures that are relevant to me. | 0.687 | 0.699 |

*Table 3: Reliability metrics for Engagement with Security Measures (SA-Engagement, alpha = .806)*

| Variable | Item | Corrected Item-Total Correlation | Alpha if Item Deleted |
|---|---|---|---|
| SAA1 | I always pay attention to experts' advice about the steps I need to take to keep my online data and accounts safe. | 0.559 | 0.722 |
| SAA2 | I am extremely knowledgeable about all the steps needed to keep my online data and accounts safe. | 0.563 | 0.727 |
| SAA3 | I am extremely motivated to take all the steps needed to keep my online data and accounts safe. | 0.675 | 0.586 |

*Table 4: Reliability metrics for Attentiveness to Security Measures (SA-Attentiveness, alpha = .762)*

| Variable | Item | Corrected Item-Total Correlation | Alpha if Item Deleted |
|---|---|---|---|
| SAR1 | I usually will not use security measures if they are inconvenient. | 0.506 | 0.743 |
| SAR2 | There are good reasons why I do not take the necessary steps to keep my online data and accounts safe. | 0.573 | 0.709 |
| SAR3 | I am too busy to put in the effort needed to change my security behaviors. | 0.612 | 0.688 |
| SAR4 | I have much bigger problems than my risk of a security breach. | 0.581 | 0.705 |

*Table 5: Reliability metrics for Resistance to Security Measures (SA-Resistance, alpha = .767)*

| Variable | Item | Corrected Item-Total Correlation | Alpha if Item Deleted |
|---|---|---|---|
| SAC1 | I want to change my security behaviors to improve my protection against threats (i.e. phishing, computer viruses, identity theft, password hacking) that are a danger to my online data and accounts. | 0.563 | 0.531 |
| SAC2 | I worry that I'm not doing enough to protect myself against threats (i.e. phishing, computer viruses, identity theft, password hacking) that are a danger to my online data and accounts. | 0.429 | 0.712 |
| SAC3 | I want to change my security behaviors in order to keep my online data and accounts safe. | 0.545 | 0.559 |

*Table 6: Reliability metrics for Concernedness with Improving Compliance (SA-Concernedness, alpha = .693)*

Lastly, we took the simple average of the sample scores for all 13 items (reversing the SA-Resistance items by subtracting each score from 6) to compute one variable that we named "SA-13," for the 13-item Security Attitude inventory.

Using Spearman correlations, which were the method used in our prior SA-6 work [30], we tested whether SA-13 would statistically associate as expected with each of its four component scales. We found the expected significant and strong positive association with the two scales that comprise SA-6, which are SA-Engagement ($r$=.720, p<.01) and SA-Attentiveness ($r$=.624, p<.01); along with a significant and moderate positive association of SA-13 with SA-Concernedness ($r$=.522, p<.01); and a significant and moderate negative association with SA-Resistance ($r$=-.461, p<.01). These tests added to our confidence that the SA-13 measurement model would hold up in subsequent analysis. Other tests of convergence and other type of validity are reported below.

## 3.3. Confirmatory Factor Analyses

Turning to Mplus, a confirmatory factor analysis again showed that all standardized loadings of the 13 indicators are above .50. Only the four-factor solution was found to have acceptable goodness-of-fit (CFI=0.933, RMSEA=0.068). By allowing three indicators to correlate, we boosted the four-factor model's CFI to .960 and reduced the RMSEA to .054. This finalized factor structure with standardized factor loadings, correlations and error terms is shown in Figure 3, with the associated fit indices summarized in Table 7.

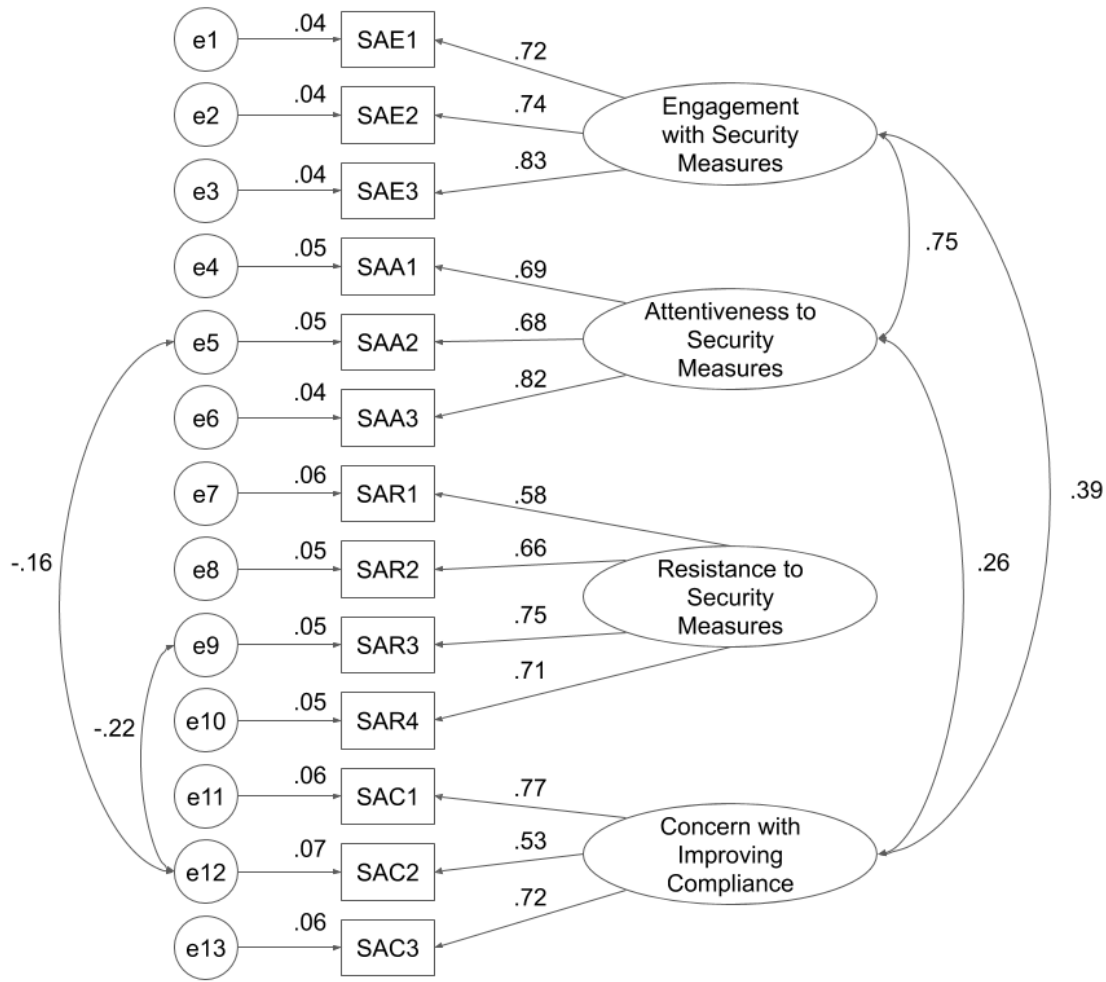

*Figure 3: Four-factor structure, standardized loadings and error terms of the SA-13 measurement model for security attitude (only statistically significant path coefficients are shown). Note that SA-Resistance is uncorrelated with the other factors. This implies that changes in someone's resistance to security measures is not related to their concern to improve compliance, and/or the degree to which they will report engaging with or paying attention to security measures.*

| N=209 | $x^2$ | df | p | $x^2/df$ | SRMR | RMSEA | TLI | CFI |
|---|---|---|---|---|---|---|---|---|
| **Null (baseline) model** | 925.983 | 78 | 0.000 | 11.872 | | | | |
| **CFA measurement model** | 115.993 | 59 | 0.000 | 1.966 | .065 | .068 | .911 | .933 |
| **Adds indicator correlations (Fig. 3)** | 89.740 | 56 | 0.003 | 1.603 | .050 | .054 | .945 | .960 |

*Table 7: Global fit indices for the SA-13 measurement model illustrated in Figure 3. The chi-statistic ratio (x2/df) below 3 indicates an adequate fit between model and data, as do the Standardized Root Mean Residual (SRMR) and Root Mean Square Error of Approximation (RMSEA) below .08, and the Tucker-Lewis (TLI) and Comparative Fit (CFI) indices above .90.*

We conducted several tests of model invariance to confirm SA-13's overall robustness [49,56,63]. To test configural invariance, we conducted two confirmatory factor analyses for group "male" (CFI=0.914, RMSEA=0.074) and group "female" (CFI=0.907, RMSEA=0.083), separately. The complete fit statistics for this and other tests of invariance [63] are reported in Table 8, with statistics indicating acceptable fit regardless of model complexity.

| Type of Invariance Test | $x^2$ | df | p | x2/df | CFI | TLI | RMSEA | SRMR | BIC | AIC | BIC/df | AIC/df |
|---|---|---|---|---|---|---|---|---|---|---|---|---|
| **CFA (baseline)** | 115.993 | 59 | 0.000 | 1.966 | 0.933 | 0.911 | 0.068 | 0.065 | 7658.490 | 7508.085 | 129.805 | 127.256 |
| **Configural invariance** | | | | | | | | | | | | |
| male | 86.307 | 59 | 0.012 | 1.463 | 0.914 | 0.887 | 0.074 | 0.081 | 3160.566 | 3051.179 | 53.569 | 51.715 |
| female | 108.162 | 59 | 0.000 | 1.833 | 0.907 | 0.878 | 0.083 | 0.077 | 4493.860 | 4367.679 | 76.167 | 74.028 |
| **Metric invariance** | 206.845 | 131 | 0.000 | 1.579 | 0.911 | 0.894 | 0.075 | 0.098 | 7661.481 | 7405.235 | 58.485 | 56.529 |
| **Intercept-only invariance** | 219.399 | 131 | 0.000 | 1.675 | 0.896 | 0.876 | 0.081 | 0.089 | 7674.036 | 7417.789 | 58.580 | 56.624 |
| **Scalar invariance** | 230.886 | 144 | 0.000 | 1.603 | 0.898 | 0.889 | 0.077 | 0.120 | 7616.260 | 7403.276 | 52.891 | 51.412 |
| **Full uniqueness invariance** | 230.886 | 144 | 0.000 | 1.603 | 0.898 | 0.889 | 0.077 | 0.120 | 7616.260 | 7403.276 | 52.891 | 51.412 |

*Table 8: CFA models and fit indices for SA-13 show acceptable fit when considering participant subsets by male and female gender. Other tests of invariance also display acceptable fit and, as a whole, indicate the CFA measurement model's robustness. Lower values for BIC/df and for AIC/df are preferred.*

# 3.4. Validity testing

## 3.4.1. Convergent validity (correlations)

A series of Spearman correlation tests found that SA-13 and its component scales showed expected statistical associations with other measures that were collected in this dataset, because they were theorized to be related to attitude or adjacent constructs in the Theory of Reasoned Action (Faklaris et al. 2019). These tests support the use of SA-13 and the SA-Engagement scale as measures of security attitude, and suggest the SA-Attentiveness, SA-Resistance and SA-Concernedness scales for specific populations, such as those who avoid making decisions, those more averse to individual risks vs. collective risks, or those exhibiting dependence or risk-taking propensities.

### 3.4.1.1. SA-13 inventory

As noted above, we found the expected significant positive association for SA-13 with the two scales that comprise SA-6, which are SA-Engagement ($r=.720$, p<.01) and SA-Attentiveness ($r=.624$, p<.01); along with a significant and moderate positive association for SA-13 with SA-Concernedness ($r=.522$, p<.01); and a significant and moderate negative association with SA-Resistance ($r=-.461$, p<.01). SA-13 was also found to be significantly positively associated with scores for SA-6 itself ($r=.744$, p<.01) and with SeBIS ($r=.537$, p<.01), the measure of security behavior intention. Finally, SA-13 was found to be significantly positively correlated with the nine-item Recalled Security Actions inventory (RSec): $r=.367$, p<.01. These tests are evidence that SA-13 can be considered a valid measure of security attitude alongside SA-6.

To make other direct comparisons with SA-6, we also conducted a number of tests with Spearman's rho, finding this boosted $r$ to .744 for SA-13's association with SA-6 and .268 for its association with SeBIS (both p<.01). We found SA-13, like SA-6, to be significantly positively associated with two measures of privacy concerns, the first of which also touches on security: the PCS ($r=.465$, p<.01) and the IUIPC ($r=.247$, p<.01). It was also significantly positively associated with Internet Know-How ($r=.494$, p<.01) and Web-oriented Digital Literacy ($r=.414$, p<.01), and with the social-behavioral and personality factors of Social Self-Efficacy ($r=.234$, p<.01) and Extraversion ($r=.178$, p<.05). These associations may indicate that SA-13 is particularly suited as a measure of security attitude in a social and/or network context.

SA-13 was found to be significantly associated with several measures for which SA-6 did not: General Decision-Making Styles subscales for avoidance ($r=.249$, p<.01) and dependence ($r$=.265, p<.01); the DoSpeRT Health/Safety measure of risk-taking propensity ($r$=.230, p<.01); and the Consideration of Future Consequences scale ($r$=-.148, p<.05). Unlike SA-6, it was not found to be associated with the Barratt Impulsiveness Scale or with measures of General Self-Efficacy, Need for Cognition or Confidence in Using Computers. These differences may indicate that SA-13 is more suited to use in populations with dependence, avoidance, or risk-taking propensities.

### 3.4.1.2. SA-Engagement scale

The SA-Engagement scale was found to be significantly positively associated with the key constructs of SA-6 ($r$=.910, p<.01), SeBIS ($r$=.441, p<.01) and RSec ($r$=.392, p<.01), along with PCS ($r$=.413, p<.01), IUIPC ($r$=.317, p<.01), Internet Know-How ($r$=.412, p<.01), Web-oriented Digital Literacy ($r$=.403, p<.01), Social Self-Efficacy ($r$=.360, p<.01) and Extraversion ($r$=.202, p<.01). Like SA-6, it was also significantly associated with the Barratt Impulsiveness Scale ($r$=-.171, p<.05), General Self-Efficacy ($r$=.156, p<.05), Need for Cognition ($r$=.232, p<.01), the DoSpeRT Health/Safety measure of risk perception ($r$=.152, p<.05), and Confidence in Using Computers ($r$=.169, p<.05). Also, like SA-6, it was not significantly associated with Consideration for Future Consequences, with the GDMS subscales for avoidance and dependence, or the DoSpeRT Health/Safety measure of risk-taking propensity.

These relationships are evidence that the SA-Engagement scale can be used as a short-form version of the SA-6 scale.

### 3.4.1.3. SA-Attentiveness scale

The SA-Attentiveness scale was found to be significantly positively associated with the key constructs of SA-6 ($r$=.895, p<.01), SeBIS ($r$=.510, p<.01) and RSec ($r$=.308, p<.01), along with PCS ($r$=.287, p<.01), IUIPC ($r$=.398, p<.01), Internet Know-How ($r$=.590, p<.01), Web-oriented Digital Literacy ($r$=.526, p<.01), and Social Self-Efficacy ($r$=.298, p<.01). Like SA-6, it was also significantly associated with the Barratt Impulsiveness Scale ($r$=-.140, p<.05), General Self-Efficacy ($r$=.224, p<.01), Need for Cognition ($r$=.232, p<.01), the DoSpeRT Health/Safety measure of risk perception ($r$=.152, p<.05), and Confidence in Using Computers ($r$=.347, p<.01). Also, like SA-6, it was not significantly associated with Consideration for Future Consequences, with the GDMS subscale for dependence, or the DoSpeRT Health/Safety measure of risk-taking propensity. Unlike the SA-Engagement scale, it was not significantly associated with Extraversion; unlike SA-6, it was significantly associated with the GDMS subscale for avoidance ($r$=.148, p<.01).

The above results show many relationships similar to SA-6 and the SA-Engagement scale, with the one significant association difference indicating that the SA-Attentiveness scale may be more suited to populations that avoid making decisions.

### 3.4.1.4. SA-Resistance scale

The SA-Resistance scale was found to be significantly associated with SeBIS ($r$=.361, p<.05), but not with SA-6 or RSec. This was expected, as this scale includes no items that are significantly associated with the items of SA-6, and its items may apply regardless of whether someone is attentive to or engaged with security advice or recalls taking specific security actions. The scale also was significantly associated with IUIPC ($r$=.253, p<.05) but not PCS; Internet Know-How ($r$=-.169, p<.05) but not Web-oriented Digital Literacy; and with Social Self-Efficacy ($r$=.148, p<.05) but not Extraversion. Its strongest significant association was with Consideration for Future Consequences ($r$=.505, p<.01), followed by GDMS-Avoidance ($r$=-.485, p<.01), General Self-Efficacy ($r$=.450, p<.01), Barratt Impulsiveness Scale ($r$=-.438,

p<.01), Confidence in Using Computers ($r$=.350, p<.01), Need for Cognition ($r$=.317, p<.01), DoSpeRT Health/Safety Risk-Taking Propensity ($r$=-.302, p<.01), and GDMS-Dependence ($r$=-.198, p<.01). The scale was not significantly associated with DoSpeRT Health/Safety Risk Perception.

These results are evidence that the SA-Resistance scale assesses different facets of security attitude than SA-6, SA-13 or the other component scales. It may be particularly suited to use in populations that are averse to individual risks, but also do not know enough about the internet to adequately perceive network or collective risks.

### 3.4.1.5. SA-Concernedness scale

The SA-Concernedness scale was found to be significantly associated with SA-6 ($r$=.193, p<.05) but not with SeBIS or RSec. This seems consistent with the item wordings, which may apply regardless of whether the respondent has formed a concrete determination to act on security advice or recalls taking specific security actions. The scale also was significantly associated with PCS ($r$=.443, p<.01) and IUIPC ($r$=.296, p<.01), as well as GDMS subscales for dependence ($r$=.302, p<.01) and avoidance ($r$=.185, p<.01), and the DoSpeRT Health/Safety measure of risk perception ($r$=.192, p<.05). It was not significantly associated with Internet Know-How, Web-oriented Digital Literacy, Social Self-Efficacy, Extraversion, General Self-Efficacy, Confidence in Using Computers, the Barratt Impulsiveness Scale, Need for Cognition, Consideration of Future Consequences, or DoSpeRT Health/Safety Risk-Taking Propensity.

These results indicate that the SA-Concernedness scale is most closely related to measures of privacy concerns. It could be useful in populations that perceive risks but depend on others to help them act in defense of these risks.

## 3.4.2. Discriminant validity (variances)

A series of independent-samples t-tests found that SA-13 and its component scales showed expected group-mean differences by amounts of security breach experiences, along with some measures of socialization (age, gender) and socio-economic status (education and income). These variables were collected with the other data because experience factors and social factors are antecedents of attitude, intention, and behavior in the Theory of Reasoned Action (Faklaris et al. 2019). The results are strong evidence of SA-13's discriminant validity and support the discriminant validity of the four component scales by some of these factors.

### 3.4.2.1. SA-13 inventory

We found that SA-13 group means varied significantly by personal experiences with security breaches (low M=3.26, SD=.47 vs. high M=3.56, SD=.50): t(101.60)=3.67, p<.001; by close ties' experiences of a security breach (low M=3.25, SD=.47 vs. high M=3.53, SD=.59): t(207)=3.84, p<.001; and by the amount heard or seen about security breaches (low M=3.21, SD=.54 vs. high M=3.42, SD=.52): t(207)=2.76, p<.01. This is strong evidence of SA-13's discriminantvalidity, as it is logical that an increase in security breach experiences will lead to an increase in someone's propensity and motivation to comply with security advice.

We also found that SA-13 group means varied significantly by gender (male M=3.46, SD=.53 vs. female or nonbinary M=3.28, SD=.53): t(207)=-2.43, p<.05; and by whether someone's yearly household income met or exceeded $25,000 (no M=3.17, SD=.51 vs. yes M=3.41, SD=.53): t(207)=2.78, p<.01. This is evidence as well of its discriminant validity, as non-male respondents may not be as socialized to have a favorable attitude toward cybersecurity evidenced by the underrepresentation of women and nonbinary employees in the field, and those with less financial means may have less motivation to protect their financial data and accounts.

In contrast with SA-6, group means did not differ significantly by age or education, suggesting that SA-13 is less sensitive to differences in security attitude related to these factors in the absence of other factors.

### 3.4.2.2. SA-Engagement

We found that the SA-Engagement scale group means differed significantly by close ties' experiences of a security breach (low M=3.49, SD=.94 vs. high M=3.76, SD=.82): t(207)=2.04, p<.05; and by the amount heard or seen about security breaches (low M=3.22, SD=1.04 vs. high M=3.77, SD=.77): t(101.69)=3.79, p<.001. This seems sufficient evidence of its discriminant validity according to experience factors. However, the scale did not vary significantly by personal experience of security breaches, suggesting that it is less sensitive to differences in these than SA-13 or SA-6 in the absence of other factors.

We also found that scale group means differed significantly by gender (male M=3.81, SD=.76 vs. female or nonbinary M=3.44, SD=.95): t(198.06)=-3.08, p<.005; by whether someone had achieved a bachelor's degree or higher (no M=3.40, SD=.97 vs. yes M=3.69, SD=.86): t(207)=2.18, p<05; and by whether someone's yearly household income met or exceeded $25,000 (no M=3.16, SD=.88 vs. yes M=3.72, SD=.87): t(207)=3.82, p<.001. The scale means did not differ significantly by age, suggesting that it is less sensitive than SA-6 to this factor.

### 3.4.2.3. SA-Attentiveness

We found that SA-Attentiveness scale group means differed significantly by the amount heard or seen about security breaches (low M=3.47, SD=.83 vs. high M=3.77, SD=.82): t(207)=2.47, p<.05, suggesting discriminant validity according to this factor. We found that SA-Attentiveness group means did not vary significantly by personal experience of security breaches, or how often a close tie had experienced a breach, suggesting this scale is less sensitive to these experiences than SA-13 or SA-6 in the absence of other factors.

This scale's group means also varied significantly by bachelor's degree or higher (no M=3.45, SD=.87 vs. yes M=3.79, SD=.80): t(207)=2.77, p<01.; and by yearly household income meeting or exceeding $25,000 (no M=3.44, SD=.87 vs. yes M=3.75, SD=.82): t(207)=2.24, p<.05, suggesting discriminant validity according to these social factors. The scale means were not significantly different by gender or age, suggesting it is less sensitive than SA-6 to these factors in the absence of other factors.

### 3.4.2.4. SA-Resistance

We found that SA-Resistance scale group means varied significantly by personal experiences with security breaches (low M=3.48, SD=.91 vs. high M=2.78, SD=.90): t(207)=-5.15, p<.001; and by how often a close tie had experienced a security breach (low M=3.43, SD=.91 vs. high M=2.97, SD=.97): t(207)=-3.42, p<.005. This suggests discriminant validity according to these experience factors. The group means did not vary significantly by how much someone had heard or seen about security breaches in the past year, suggesting that SA-Resistance is not sensitive to this factor in the absence of other factors.

The scale group means also differed significantly by gender (male M=3.09, SD=.99 vs. female or nonbinary M=3.38, SD=.92): t(207)=2.17, p<.05; and by whether someone was younger than 40 (M=2.98, SD=.89) vs. 40 or older (M=3.44, SD=.97): t(207)=35.11, p<.005. This seems sufficient evidence of its discriminant validity according to social factors. The scale group means were not significantly different by college attainment or income level, suggesting that SA-Resistance is not sensitive to these factors in the absence of other factors.

### 3.4.2.5. SA-Concernedness

We found that SA-Concernedness scale group means varied significantly by personal experiences with security breaches (low M=3.52, SD=.86 vs. high M=3.80, SD=.78): t(207)=2.16, p<.05; and by how often a close tie had experienced a security breach (low M=3.48, SD=.91 vs. high M=3.82, SD=.66): t(207)=2.91, p<.01. This seems sufficient evidence of its discriminant validity according to social factors. The scale group means did not vary significantly by how much someone had heard or seen about security breaches in the past year, suggesting that SA-Concernedness is not sensitive to this factor in the absence of other factors.

This scale's group means also differed significantly by whether someone's yearly household income exceeded $25,000 (no M=3.27, SD=.85 vs. yes M=3.70, SD=.82): t(207)=3.16, p<.005. This supports discriminant validity according to this factor. The scale group means did not differ significantly by gender, age or education, suggesting that SA-Concernedness is not sensitive to these factors in the absence of other factors.

## 3.5. Predictive validity (regressions)

Finally, we examined how well SA-13 and its component scales fit into models based on the Theory of Reasoned Action (TRA) applied to end-user cybersecurity (Faklaris et al. 2019). We found excellent fit particularly for the regression models, supporting the overall validity and usefulness of SA-13 and its components scales as measures of attitude that can help predict and explain security behavior intention and security behavior.

### 3.5.1. Theory of Reasoned Action (TRA) and SA-13 predictiveness

In the TRA, attitude is an antecedent of behavior intention and of behavior [3–5,32]. Extending this to end-user cybersecurity, Faklaris et al. (2019) found both a direct effect for SA-6 (the previous security attitude measure) on RSec (the behavior measure), and an indirect effect mediated by SeBIS (the intention measure). Guided by this updated model for cybersecurity (Figure 4), we tested the effect of SA-13 and the component scales on RSec using linear and stepwise regression (Table 9) to determine which of these relationships are significantly reliant on SeBIS' mediation. Results are summarized below.

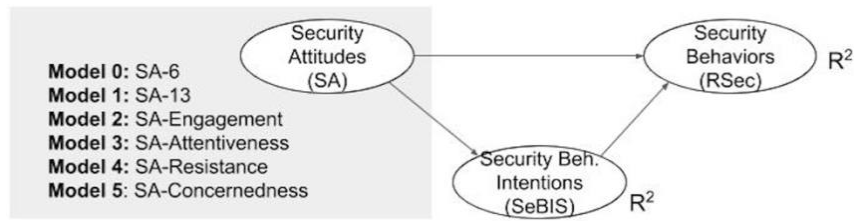

*Figure 4: Attitude has both a direct effect on behavior as well as an indirect effect mediated by behavior intention, according to the Theory of Reasoned Action for end-user cybersecurity (Faklaris et al. 2019). In models 0-5, we test different measures for attitude (SA) as predictors of intention (SeBIS, Egelman and Peer 2015) and behavior (RSec, Faklaris et al. 2019). The amount of variance explained in the two outcome variables by the various predictor combinations is represented by $R^2$.*

| Model | SeBIS $R^2$ | ß | p | Step | RSec Predictors added | $R^2$ | Change in $R^2$ | Step 1 ß | p | Step 2 ß | p |
|---|---|---|---|---|---|---|---|---|---|---|---|
| **Model 0** | | | | Step 1 | SA-6 | .131**** | | .362**** | .000 | .391**** | .000 |
| | .280**** | .529**** | .000 | Step 2 | SeBIS | .133 | .002 | | | -.054 | .483 |
| **Model 1** | | | | Step 1 | SA-13 | .104**** | | .322**** | .000 | .303**** | .000 |
| | .046*** | .215*** | .002 | Step 2 | SeBIS | .111 | .007 | | | .088 | .193 |
| **Model 2** | | | | Step 1 | SA-Engagement | .129**** | | .359**** | .000 | .362**** | .000 |
| | .194**** | .440**** | .000 | Step 2 | SeBIS | .129 | .000 | | | -.007 | .928 |
| **Model 3** | | | | Step 1 | SA-Attentiveness | .082**** | | .287**** | .000 | .282**** | .000 |
| | .259**** | .509**** | .000 | Step 2 | SeBIS | .082 | .000 | | | .009 | .904 |
| **Model 4** | | | | Step 1 | SA-Resistance | .013** | | -.113 | .104 | -.187** | .010 |
| | .116**** | .341**** | .000 | Step 2 | SeBIS | .054*** | .041*** | | | .217*** | .003 |
| **Model 5** | | | | Step 1 | SA-Concernedness | .002 | | .045 | .520 | .025 | .715 |
| | .017* | .129* | .063 | Step 2 | SeBIS | .024* | .022* | | | .150* | .032 |

*Table 9: Adding to evidence for their validity, all but one of the scales (SA-Concernedness) had a direct estimated effect on scores for both SeBIS and for RSec that was significantly different from zero (alpha = .05), as posited by*

*the Theory of Reasoned Action for end-user cybersecurity (Faklaris et. al 2019). Standardized beta coefficients show the direction and strength of the relationship:  +p<.10, \*p<.05, \*\*p<.01, \*\*\*p<.005, \*\*\*\*p<.001; effect sizes: 0.02= small, 0.15=medium, 0.35=large.*

At alpha = .05, we found that all but one of the scales (SA-Concernedness) had a direct effect on scores for both SeBIS and for RSec that was significantly different from zero. This adds evidence for their validity as attitude measurements according to the TRA, which posits that attitude is an antecedent of intention and behavior.

### 3.5.2. SeBIS not needed as a predictor of RSec

Moreover, the effect of adding SeBIS as a predictor of RSec in models 0-3 was *not* significantly different from zero. This suggests that SA-6, SA-13, SA-Engagement and SA-Attentiveness can be used in place of SeBIS to directly predict or explain security behaviors.

### 3.5.3. SA-13 subscales as diagnostics of security behavior change

Finally, the pattern of significant $R^2$ values for RSec suggest that the component scales help explain a continuum of security behavior change. Specifically, SA-Engagement alone accounts for 12.9% of the variance in RSec; SA-Attentiveness alone, 8.2%; SA-Resistance mediated by SeBIS, 5.4%; and SA-Concernedness mediated by SeBIS, 2.4%. This represents a medium effect size for SA-Engagement (Cohen's $f^2$=.15) and a small effect size for SA-Attentiveness (Cohen's $f^2$=.09), for SA-Resistance mediated by SeBIS (Cohen's $f^2$=.04), and for SA-Concernedness mediated by SeBIS (Cohen's $f^2$=.02).

# 4. Discussion

Our methods and results showed that, with the SA-13 inventory (Tables 1-2, Appendix E), we have developed a reliable and valid measurement for security attitudes that addresses cybersecurity decisional balance. This four-factor model (Figure 3) displayed desirable goodness-of-fit, measurement invariance, and convergent, discriminant and predictive validity when compared with other variables in the dataset. Further, of the resulting four subscales — *Engagement with Security Measures (SA-Engagement* for short), *Attentiveness to Security Measures (SA-Attentiveness), Resistance to Security Measures (SA-Resistance)* and *Concernedness with Improving Compliance (SA-Concernedness)* – we found the first three (Tables 3-5) displayed desirable internal consistency and convergent, discriminant and predictive validity, supporting their use as independent scales.

## 4.1. SA-13 and Decisional Balance

Cybersecurity psychology is akin to health psychology in that, in the short-term, people often act against their long-term or collective interest by neglecting to take precautions or avoid risky behaviors [29,70]. From the user's point of view, this is not necessarily irrational, as they may have needs that conflict with security recommendations and policies [1]; also, people's decision-making styles vary [65], which impacts their intentions and behaviors in both health care [31,50] and cybersecurity [27,54]. Indeed, several studies in usable security and human-computer interaction [16,35,62,74] have noted users resisting security tools or best practices as part of engaging in a mental process of weighing the "pros and cons" [45]. However, until now, few security researchers have explicitly looked for insights in the literature on decisional balance and health behavior change.

Our SA-13 inventory provides a reliable and valid measurement of a person's security attitudes with respect to this decisional balance. It expands the original SA-6 scale to include seven items that address security noncompliance, three of which cover someone's *concernedness* about this noncompliance, and four which assess the degree of bad feeling or *resistance* to these requirements. These weigh in balance with the original six items, which address *engagement* with and *attentiveness* to these requirements, to compute a holistic score of someone's security attitudes. Thus, SA-13 has potential to act as a diagnostic for a person's cybersecurity decision-making in studies of security behavior change, similar to how decisional balance inventories have been used for assessing elements of health behavior change [20,55,57,58,71].

Further, our SA-13 inventory makes it possible to investigate mental models around security without the expenses or logistical challenges of interviews, but while still capturing the resistance and concernedness that participants might be thinking or feeling with regards to security compliance. SA-13 facilitates making useful comparisons via scores, such as which roommates in a household group have security attitudes that are the most or least positive, and the degree to which that is associated with the group's overall security compliance. SA-13 would be particularly useful in conducting user studies at scale, such as in workplaces concerned with how best to direct their budget for security awareness training or in research across a social media platform to determine which subgroups of users may be more vulnerable to a social engineering attack. In such situations, SA-13 can help document the impact of security interventions when teams are unable, for confidentiality or privacy reasons, to directly access users' system log data. It is suited to be delivered via paper, desktop computer, or mobile device.

Finally, our results show SA-Resistance itself is a reliable and valid measure and can be used as an independent scale. This will help industry professionals and researchers to assess whether a given set of employees or research participants are open to trying out a newly developed tool or to pilot-testing a new set of security procedures. It also might prove helpful in evaluating the amount of "hassle" that a person thinks is involved in cybersecurity and how likely they are to bypass security measures, as well as how this interacts with security behavior intention and actual security behavior in the given system.

## 4.2. When to use SA-Engagement vs. SA-Attentiveness

Of the two SA-13 factors representing security compliance, originally developed as SA-6 [30], we see slightly different contexts in which an industry professional or academic researcher might employ one or the other as independent scales.

SA-Engagement seems the best suited to act as a shorter version of SA-6 to measure security compliance. This scale displayed similar significant associations as SA-6 with other measures in the dataset, and a nearly $r=1.00$ association with SA-6 itself. Substituting SA-Engagement for SA-6 would be useful in a survey protocol or in the screen interface for an online study that already is stuffed with other questions, as every item added increases the respondents' fatigue and likelihood to abandon the study or to stop answering honestly. Also, regressions suggest SA-Engagement is better suited to use in studies that seek to directly predict and explain security behavior, as it significantly accounted for 4.7% more variance in RSec, the behavior measure, than did SA-Attentiveness (Table 9).

Meanwhile, SA-Attentiveness, which also displays desirable psychometric properties, seems to correspond most closely to the definition of security sensitivity [15,16], as its three items rate someone's degree of awareness, motivation and knowledge to use security tools and advice. This points to its use as most appropriate in studies that are grounded in this security sensitivity work or similar research, but likewise are concerned to limit the length of survey protocols and prevent respondent fatigue and abandonment. Also, regressions suggest SA-Attentiveness is better suited to use in studies that seek to directly predict and explain security behavior intention, as it significantly accounted for 6.5% more variance in SeBIS, the behavior intention measure, than did SA-Engagement (Table 9).

## 4.3. Limitations and Future Work

Our study re-analyzes data that was collected for the creation of the SA-6 security attitude measure [30], which allows us to draw direct comparisons with those results and to build on their careful method of psychometric scale development without the expense of recruiting a new, U.S. Census-weighted survey panel. A future effort to collect survey data specifically focused on security noncompliance may be able to generate items for SA-Concernedness, in particular, that will exhibit better psychometric properties. We note that this data comes from U.S.-based adult participants that were recruited with purposive, nonrandom sampling, which limits the ability to generalize results inside the U.S. or to teenagers. Our use, in the prior SA-6 work, of a long online questionnaire in English also may have introduced language bias and/or common method bias. More research will be needed to support that SA-13 is reliable and valid when administered in other languages or modes, or in countries outside of the U.S.

The regression results for SA-13, while unable to prove causation alone, point to several directions for future research into users' cybersecurity decision-making. The many statistically significant path coefficients suggest that SA-6, SA-13, SA-Engagement and SA-Attentiveness can be used in place of SeBIS to directly predict or explain security behaviors, which experimental studies should investigate further. With regards to SA-Resistance, the stepwise regressions (Table 9, Model 4) suggest that participants with higher levels of SA-Resistance are still highly likely to intend to act. More research is needed to determine what might explain these results, such as the degree to which any background factors can significantly impact a person's level of security resistance and its relationship with their behavior intention and behavior.

# 5. Conclusion

In this study, we have described the development and validation of SA-13, a new 13-item survey measure of security attitude that balances self-ratings of both compliance and noncompliance with cybersecurity measures. We showed that three of SA-13's four subscales — *Engagement with Security Measures* (SA-Engagement, three items), *Attentiveness to Security Measures* (SA-Attentiveness, three items), and *Resistance to Security Measures* (SA-Resistance, four items) — can be used independently as reliable and valid measures. Further, regressions suggest that SA-6, SA-13, SA-Engagement and SA-Attentiveness all can be used in place of SeBIS, the current state-of-the-art measure of security behavior intention, to account for variances in RSec, the proxy measure of security behavior. Lastly, we discussed SA-13's potential to act as a diagnostic for a person's cybersecurity decision-making in studies of security behavior change, similar to how inventories are used in health research to assess elements of decisional balance.

Our study provides researchers and industry professionals with valuable tools beyond system log data with which to measure computer users' acceptance and resistance to complying with security policies and following security advice. We hope this will help the fields of psychology, human-computer interaction and cybersecurity to explain, predict and influence the behaviors of computer users in ways that head off social engineering attacks and improve the security of computer networks.

# Acknowledgements

This work was generously supported by the U.S. National Science Foundation, grant no. CNS-1704087. The first author also is grateful for fellowship support from the CyLab Security and Privacy Institute and the Center for Informed Democracy and Social Cybersecurity, both at Carnegie Mellon University, and for feedback on previous versions of this work by reviewers for the USENIX Symposium on Usable Privacy and Security (SOUPS), by the Workshop on Security Information Workers (WSIW), by members of the Connected Experiences Lab at the HCII, and many others. Sponsors were not involved in any phase of research or article preparation.

# Appendix A

The following items were tested for possible inclusion in the multi-factor attitude model. Those in bold comprise the

Security Attitude inventory (SA-13) and scales as documented in this article.

*Table 10: Candidate items for SA-13 and their sources*

| Candidate items (n=48) analyzed for scale | Source |
|---|---|
| A security breach, if one occurs, is not likely to cause significant harm to my online identity or accounts. | [1,15,16,18] |
| Generally, I am aware of existing security threats. | [15–18] |
| Generally, I am willing to spend money to use security measures that counteract the threats that are relevant to me. | [16,40] |
| Generally, I care about security and privacy threats. | [15–18] |
| **Generally, I diligently follow a routine about security practices.** | [30] |
| Generally, I know how to figure out if an email was sent by a scam artist. | [15] |
| Generally, I know how to use security measures to counteract the threats that are relevant to me. | [15–18] |
| Generally, I know which security threats are relevant to me. | [15–18] |
| Generally, I want to use measures that can counteract security and privacy threats. | [15–18] |
| **I always pay attention to experts' advice about the steps I need to take to keep my online data and accounts safe.** | [14,16] |
| I always trust experts' recommendations about security measures (such as using unique passwords or a password manager, installing recommended software updates, etc.). | [14,16] |
| I am confident that I am taking the necessary steps to keep my online data and accounts safe. | [15–18] |
| I am confident that I can change my security behaviors, if needed, to protect myself against threats (such as phishing, computer viruses, identity theft, password hacking) that are a danger to my online data and accounts. | [77] |
| I am confident that I could change my security behaviors if I decided to. | [77] |
| **I am extremely knowledgeable about all the steps needed to keep my online data and accounts safe.** | [15–18] |
| I am extremely knowledgeable about how to take the necessary steps to keep my online data and accounts safe. | [15–18] |
| I am extremely knowledgeable about which security threats (such as phishing, computer viruses, malware, password hacking) are a danger to my online data and accounts. | [15–18] |
| **I am extremely motivated to take all the steps needed to keep my online data and accounts safe.** | [15–18] |
| I am extremely well aware of existing security threats (such as phishing, computer viruses, identity theft, password hacking). | [15–18] |
| I am extremely well aware of the necessary steps that I can take to counteract security threats (such as phishing, computer viruses, identity theft, password hacking). | [15–18] |
| **I am too busy to put in the effort needed to change my security behaviors.** | [16,22] |
| I care very much about the issue of security threats (such as phishing, computer viruses, identity theft, password hacking). | [15–18] |
| I dread that using recommended security measures will backfire on me (such as forgetting a needed password, updated software becoming unusable, etc.). | [16,40] |
| I feel guilty when I do not use recommended security measures (such as by reusing passwords, putting off software updates, etc.). | [16] |
| I generally am aware of existing security measures that I can use to counteract security threats. | [15–18] |
| I generally am aware of methods to send email or text messages that can't be spied on. | [15–18] |
| **I have much bigger problems than my risk of a security breach.** | [16,22] |
| I need to change my security behaviors to improve my protection against security threats (such as phishing, computer viruses, identity theft, password hacking). | [15,77] |
| **I often am interested in articles about security threats.** | [19] |
| **I seek out opportunities to learn about security measures that are relevant to me.** | [16] |
| **I usually will not use security measures if they are inconvenient.** | [15–18] |
| I usually will not use security measures unless I am forced to. | [15–18] |
| **I want to change my security behaviors in order to keep my online data and accounts safe.** | [15,77] |
| **I want to change my security behaviors to improve my protection against threats (such as phishing, computer viruses, identity theft, password hacking) that are a danger to my online data and accounts.** | [15,77] |
| **I worry that I'm not doing enough to protect myself against threats (such as phishing, computer viruses, identity theft, password hacking) that are a danger to my online data and accounts.** | [15,60] |
| It is a lost cause to take all the steps needed to keep my online data and accounts safe. | [30] |
| It is important for me to change my security behaviors to improve my protection against security threats (such as phishing, computer viruses, identity theft, password hacking). | [15,77] |
| It is not possible for me to do more than I already am to counteract security threats (such as phishing, computer viruses, identity theft, password hacking) that are a danger to my online data and accounts. | [30] |
| It's a sign of paranoia to use numerous security measures to protect against threats. | [16,35] |
| It's a sign of paranoia to use recommended security measures (such as using unique passwords or a password manager, installing recommended software updates, etc.). | [16,35] |
| My current lapses in using security measures are harmless. | [1,16] |
| My own actions can make a significant difference in keeping my online data and accounts safe. | [10] |
| Oftentimes, as soon as I discover a security problem, I report it to someone who can fix it. | [25] |
| Oftentimes, I am running on "automatic pilot" when I sift through my email and text messages. | [30] |
| Oftentimes, I will check that my anti-virus software has been regularly updating itself. | [25] |
| The exposure of my online data and accounts in a security incident, if one occurs, would be a significant problem for me. | [30] |
| The theft of my online data or accounts in a security breach, if one occurs, would be a significant problem for me. | [30] |
| **There are good reasons why I do not take the necessary steps to keep my online data and accounts safe.** | [16] |

# Appendix B

The following table presents the means, standard deviations, and correlations among the 13

survey items in the online survey responses used for exploratory factor analysis ($N_{EFA}$=475).

| Variable | Item | Correlations for EFA Dataset (N=678, * p<.05, ** p<.01) |||||||||||||
|---|---|---|---|---|---|---|---|---|---|---|---|---|---|---|
| | | SAE1 | SAE2 | SAE3 | SAA1 | SAA2 | SAA3 | SAR1 | SAR2 | SAR3 | SAR4 | SAC1 | SAC2 | SAC3 |
| SAE1 | I often am interested in articles about security threats. | 1 | | | | | | | | | | | | |
| SAE2 | Generally, I diligently follow a routine about security practices. | .541** | 1 | | | | | | | | | | | |
| SAE3 | I seek out opportunities to learn about security measures that are relevant to me. | .607** | .599** | 1 | | | | | | | | | | |
| SAA1 | I always pay attention to experts' advice about the steps I need to take to keep my online data and accounts safe. | .336** | .405** | .491** | 1 | | | | | | | | | |
| SAA2 | I am extremely knowledgeable about all the steps needed to keep my online data and accounts safe. | .356** | .442** | .441** | .424** | 1 | | | | | | | | |
| SAA3 | I am extremely motivated to take all the steps needed to keep my online data and accounts safe. | .421** | .463** | .463** | .574** | .568** | 1 | | | | | | | |
| SAR1 | I usually will not use security measures if they are inconvenient. | .000 | .132 | .022 | .096 | -.005 | .143* | 1 | | | | | | |
| SAR2 | There are good reasons why I do not take the necessary steps to keep my online data and accounts safe. | -.002 | -.018 | -.095 | -.129 | -.176* | -.056 | .439** | 1 | | | | | |
| SAR3 | I am too busy to put in the effort needed to change my security behaviors. | -.041 | .024 | -.014 | .074 | -.019 | .144* | .423** | .468** | 1 | | | | |
| SAR4 | I have much bigger problems than my risk of a security breach. | .017 | .056 | .029 | .018 | -.110 | .058 | .371** | .460** | .554** | 1 | | | |
| SAC1 | I want to change my security behaviors to improve my protection against threats (i.e. phishing, computer viruses, identity theft, password hacking) that are a danger to my online data and accounts. | .210** | .172* | .304** | .106 | .032 | .180** | .005 | -.087 | -.067 | .017 | 1 | | |
| SAC2 | I worry that I'm not doing enough to protect myself against threats (i.e. phishing, computer viruses, identity theft, password hacking) that are a danger to my online data and accounts. | .064 | -.055 | .093 | -.081 | -.149* | -.084 | -.052 | -.102 | -.265** | -.045 | .391** | 1 | |
| SAC3 | I want to change my security behaviors in order to keep my online data and accounts safe. | .210** | .199** | .205** | .179** | .081 | .231** | .086 | .029 | -.070 | .002 | .553** | .366** | 1 |
| | Mean (1.00-5.00 scale) | 3.50 | 3.74 | 3.53 | 3.79 | 3.44 | 3.80 | 3.12 | 3.22 | 3.52 | 3.18 | 3.72 | 3.37 | 3.72 |
| | Standard Deviation | 1.12 | 0.99 | 1.07 | 0.91 | 1.12 | 1.01 | 1.24 | 1.32 | 1.22 | 1.20 | 1.04 | 1.16 | 1.00 |

*Table 11: Adding to evidence for their validity, all but one of the scales (SA-Concernedness) had a direct estimated effect on scores for both SeBIS and for RSec that was significantly different from zero (alpha = .05), as posited by the Theory of Reasoned Action for end-user cybersecurity (Faklaris et. al 2019). Standardized beta coefficients show the direction and strength of the relationship: +p<.10, *p<.05, **p<.01, ***p<.005, ****p<.001; effect sizes: 0.02= small, 0.15=medium, 0.35=large.*

# Appendix C

The following table presents the factor loadings and matrix for the unrotated four-factor solution in the $N_{\text{EFA}}$=475 dataset. The scree plot below the table shows a flattening after Factor 4, indicating that Factors 5-13 are primarily measuring error variance.

*Table 12: Factor solution for SA-13 in the exploratory dataset*

| Variable | Item | Factor | | | |
|---|---|---|---|---|---|
| | | 1 | 2 | 3 | 4 |
| SAA4 | I always pay attention to experts' advice about the steps I need to take to keep my online data and accounts safe. | 0.779 | -0.128 | -0.139 | 0.121 |
| SAE3 | I seek out opportunities to learn about security measures that are relevant to me. | 0.771 | -0.167 | 0.016 | -0.239 |
| SAA3 | I am extremely motivated to take all the steps needed to keep my online data and accounts safe. | 0.763 | -0.015 | -0.013 | 0.085 |
| SAE2 | Generally, I diligently follow a routine about security practices. | 0.749 | -0.111 | -0.165 | 0.114 |
| SAE1 | I often am interested in articles about security threats. | 0.746 | -0.223 | 0.043 | -0.284 |
| SAA2 | I am extremely knowledgeable about all the steps needed to keep my online data and accounts safe. | 0.637 | -0.237 | -0.236 | 0.227 |
| SAR4 | I have much bigger problems than my risk of a security breach. | 0.246 | 0.655 | -0.231 | -0.166 |
| SAR2 | There are good reasons why I do not take the necessary steps to keep my online data and accounts safe. | 0.066 | 0.652 | -0.181 | 0.082 |
| SAR3 | I am too busy to put in the effort needed to change my security behaviors. | 0.331 | 0.574 | -0.376 | -0.025 |
| SAR1 | I usually will not use security measures if they are inconvenient. | 0.034 | 0.358 | -0.128 | 0.03 |
| SAC1 | I want to change my security behaviors to improve my protection against threats (i.e. phishing, computer viruses, identity theft, password hacking) that are a danger to my online data and accounts. | 0.353 | 0.276 | 0.665 | 0.085 |
| SAC3 | I want to change my security behaviors in order to keep my online data and accounts safe. | 0.365 | 0.302 | 0.618 | 0.083 |
| SAC2 | I worry that I'm not doing enough to protect myself against threats (i.e. phishing, computer viruses, identity theft, password hacking) that are a danger to my online data and accounts. | 0.169 | 0.159 | 0.507 | -0.06 |

Extraction method: Principal Axis Factoring. Rotation method: None.

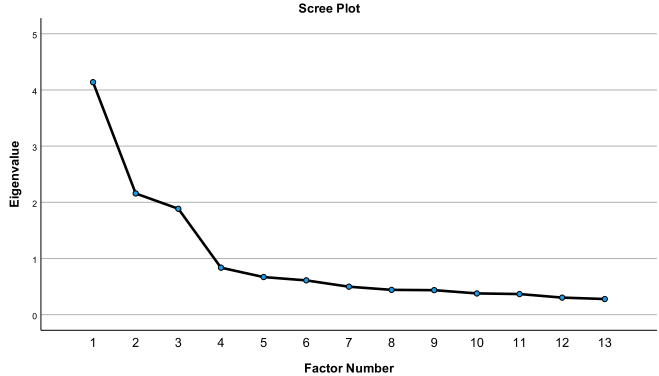

*Figure 5: Scree plot for the SA-13 factor solution in the exploratory dataset*

# Appendix D

Cronbach's alpha (threshold = .70) for composite measures included for convergent and discriminant validity testing in the Qualtrics dataset (*N*=209).

| Measure | Alpha |
|---|---|
| Barratt Impulsivity Scale | 0.86 |
| Big5-Agreeableness | 0.40 |
| Big5-Conscientiousness | 0.53 |
| Big5-Emotional Stability | 0.60 |
| Big5-Extraversion | 0.70 |
| Big5-Openness to Experiences | 0.37 |
| Confidence in Using Computers | 0.89 |
| Consideration of Future Consequences | 0.77 |
| DoSpERT - Risk perception subscale | 0.89 |
| DoSpERT - Risk-taking subscale | 0.84 |
| GDMS – Avoidance subscale | 0.91 |
| GDMS – Dependence subscale | 0.81 |
| Kang Internet Know-How scale | 0.91 |
| Kang Technical Know-How scale | 0.63 |
| Need for Cognition scale | 0.88 |
| Privacy – Internet Users' Infor. Privacy Concerns | 0.88 |
| Privacy – Privacy Concerns Scale | 0.96 |
| SeBIS – Security Behavior Intentions Scale | 0.70 |
| Self-Efficacy - General | 0.90 |
| Self-Efficacy - Social | 0.75 |
| Web-oriented Digital Literacy | 0.94 |

*Table 13: Measures collected in the SA-6 dataset and their Cronbach's alpha scores.*

# Appendix E

Below, we provide a summary of how to administer and score SA-13 and its subscales.

## Directions to give to participants:

Each statement below describes how a person might feel about the use of security measures. Examples of security measures are laptop or tablet passwords, spam email reporting tools, software updates, secure web browsers, fingerprint ID, and anti-virus software.

Please indicate the degree to which you agree or disagree with each statement. In each case, make your choice in terms of how you feel **right now**, not what you have felt in the past or would like to feel.

There are no wrong answers.

## Response set:

1=Strongly disagree, 2=Somewhat disagree, 3=Neither disagree nor agree, 4=Somewhat agree, 5=Strongly agree

## Items (randomize order if possible):

1. I seek out opportunities to learn about security measures that are relevant to me.

2. I am extremely motivated to take all the steps needed to keep my online data and accounts safe.

3. Generally, I diligently follow a routine for security practices.

4. I often am interested in articles about security threats.

5. I always pay attention to experts' advice about the steps I need to take to keep my online data and accounts safe.

6. I am extremely knowledgeable about all the steps needed to keep my online data and accounts safe.

7. I am too busy to put in the effort needed to change my security behaviors. *(reverse for computing full scale)*

8. I have much bigger problems than my risk of a security breach. *(reverse for computing full scale)*

9. There are good reasons why I do not take the necessary steps to keep my online data and accounts safe. *(reverse for computing full scale)*

10. I usually will not use security measures if they are inconvenient. *(reverse for computing full scale)*

11. I want to change my security behaviors to improve my protection against threats (*e.g.* phishing, computer viruses, identity theft, password hacking) that are a danger to my online data and accounts.

12. I want to change my security behaviors in order to keep my online data and accounts safe.

13. I worry that I'm not doing enough to protect myself against threats (*e.g.* phishing, computer viruses, identity theft, password hacking) that are a danger to my online data and accounts.

## Scoring:

- *Engagement* subscale (SA-6): Mean of items 1, 3, 4

- *Attentiveness* subscale (SA-6): Mean of items 2, 5, 6

- *Resistance* subscale: Mean of items 7-10

- *Concernedness* subscale*:* Mean of items 11-13

- *Overall:* Reverse the Resistance items (recode responses as 6-*r*), then take the mean of all 13 items.